\documentclass[reprint,showpacs,preprintnumbers, amsmath,amssymb,aps,prc,]{revtex4-1}
\usepackage{amsfonts}
\usepackage{amsmath}
\usepackage{amssymb}
\usepackage{graphicx,color}
\usepackage{rotating}
\usepackage{hyperref}
\usepackage{dcolumn}
\usepackage{bm,ulem}

\begin{document}

\title{Repulsive Vector Interaction in Three Flavor Magnetized Quark and Stellar Matter}

\author{D\'{e}bora P. Menezes}
\author{Marcus B. Pinto}
\author{Luis B.  Castro}
\affiliation{Depto de F\'{\i}sica - CFM - Universidade Federal de Santa
Catarina  Florian\'opolis - SC - CP. 476 - CEP 88.040 - 900 - Brazil}
\author{Pedro Costa}
\author{Constan\c ca Provid\^encia}
\affiliation{Centro de F\'{\i}sica Computacional - Department of Physics -
University of Coimbra - P-3004 - 516 - Coimbra - Portugal}

\begin{abstract}
The effect of the vector interaction on  three flavor magnetized matter is
studied within the SU(3) Nambu--Jona-Lasiono quark model. We have considered 
cold
matter under a static external magnetic field within two different models for 
the vector interaction in order to investigate how the
form of the vector interaction and the intensity of the magnetic field affect
the equation of state as well as  the strangeness content. It was shown that the
flavor independent vector interaction predicts a smaller strangeness content 
and, therefore, harder equations of state. On the other hand, the flavor
dependent vector interaction favors larger strangeness content the
larger the vector coupling.  We have confirmed that at low
densities the magnetic field and the vector interaction have opposite
competing effects: the first one softens the equation of state while
the second hardens it. Quark stars and hybrid stars
subject to an external magnetic field were also studied. Larger star masses
are obtained for the flavor independent vector interaction. Hybrid
stars may bare a core containing deconfined quarks if neither the vector
interaction nor the magnetic field are  too strong. Also, the presence of 
strong magnetic fields seems to disfavor the existence of a quark core in
hybrid stars. 

\end{abstract}

\pacs{12.39.Ki,24.10.Jv,26.60.Kp}

\maketitle

\section{Introduction} 

Early investigations performed with the Walecka model for nuclear matter 
\cite{walecka} show that the inclusion of a vector-isoscalar channel is an
essential ingredient for  an accurate description of nuclear matter. Later, such a channel has been considered to extend the standard Nambu--Jona-Lasinio model (NJL), which originally included only a scalar and a pseudoscalar type of channels, in order to obtain a saturating chiral theory for nuclear matter described only by fermions \cite {volker}. As discussed in Ref. \cite{ruggieri2009} 
the introduction of the vector interaction, and thus of the vector excitations, is also important in determining the properties of strongly interacting matter at intermediate 
densities where  vector mesons mediate the interactions and their exchange might be responsible 
for kaon condensation at high density.   Recently, the presence of a vector interaction in the NJL model  was crucial to reproduce the measured relative elliptic flow differences between nucleons and 
anti-nucleons as well as between kaons and antikaons at energies 
carried out in the Beam-Energy Scan 
program of the Relativistic Heavy Ion Collider \cite {ko}.   
 
Regarding the QCD phase diagram at finite quark density it has been established that the net effect of a repulsive vector contribution  is  to weaken the first order transition \cite {fuku08}. Indeed, it has been observed that the first order transition region shrinks, forcing the critical end point (CEP) to appear at smaller temperatures,  while the first order transition occurs at higher  
chemical potential values when the vector interaction increases. 

Since the finite density region of the QCD phase diagram is not yet accessible to lattice simulations one usually employs model approximations to study the associated phase transitions as well as to evaluate the equation of state  (EOS) to be used in stellar modeling. One of the most popular models adopted in these investigations is the NJL which, as already referred,  can be easily extend to accommodate a vector channel while keeping the original symmetries. 
At present, despite its importance, the vector term  coupling  $G_V$ cannot 
be determined from experiments and lattice QCD simulations, although
there have been some attempts to determine its value. For instance, 
in Ref. \cite{klimt} a vector coupling constant of the order of magnitude of 
the scalar--pseudoscalar coupling was obtained by fitting the nucleon axial 
charge or masses of vector mesons and in \cite{hanauske}, the pion mass and 
the pion decay constant were recalculated as a function of the vector 
interaction and shown to vary by about 10\% when for $0 < x < 1$, where 
$x=G_V/G_S$, $G_S$ being the scalar coupling.
Eventually, the combination of neutron star observations and the energy scan 
of the  phase-transition signals 
at FAIR/NICA may provide us some hints on the precise numerical value. Meanwhile, $G_V$ has been 
taken as a free parameter in most works.
Finally, note that this channel interaction can be generated by higher order (exchange type of contributions) which are present in approximations which go beyond the large-$N_c$ limit like the  the nonperturbative Optimized Perturbation Theory (OPT) \cite {prcopt}.  

The fact that a strong enough vector term may turn the first order phase 
transition, which is expected at the low temperature part of the QCD phase diagram, into a smooth cross over (for the realistic case of quarks with finite 
current masses) may also have  astrophysical implications affecting  the 
structure
of the compact stellar objects. In Ref.  \cite{hanauske} a variable vector coupling was used in the 
discussion of the possible properties of quark stars  and
the authors have shown that depending on the value of the vector coupling  
the star could either be self-bound and present a finite density at the 
surface or bear a very small density at the surface, behaving as a 
standard (hadronic) neutron star. The maximum stellar mass obtained, 
$M=1.6\, M_\odot$, corresponds to the largest vector coupling considered, 
$x=1$, i.e., $G_V=G_S$. After these seminal works, a
repulsive vector term was also used in many other investigations involving 
hybrid stars and possible phase transitions to a quark phase \cite{pagliara}.
Recently, the importance of the vector interaction in describing massive 
stars has also been extensively discussed \cite{bonanno,lenzi2012,logoteta,shao2013,sasaki2013,masuda2013}. 

Another timely important problem concerns the  investigation of the effects 
produced by a  magnetic field $B$ on the QCD phase diagram and also on the EOS 
used to model neutron stars. The motivation stems from the fact that strong magnetic fields may be produced  in non central heavy ion collisions \cite {kharzeev09,tuchin}, as well as being present in magnetars \cite {magnetars}.

Regarding stellar matter the low temperature part of the QCD phase diagram, where a  first order (chiral) phase transition is expected to occur \cite {scavenius, prcopt}, constitutes the relevant region to be investigated.
The question of how this region is 
affected by magnetic fields has been addressed in Refs. \cite {prd} and 
\cite{pedro2013} in the framework of the three flavor NJL and PNJL models 
respectively. 
One of the main results of Ref. \cite {prd}  shows that in this regime  the symmetry broken phase tends to shrink  with increasing values of $B$. At these low temperatures, the chemical potential value associated with the first order transition decreases with increasing magnetic fields, effect known as the inverse magnetic catalysis phenomenon (IMC). This result has been previously observed with the two flavor NJL, in the chiral limit  \cite {inagaki}, as well as with  a holographic one-flavor model \cite {andreas} and more recently with the planar Gross-Neveu model \cite {novo}.  A model-independent physical explanation for IMC is given  in Ref. \cite {andreas} while a recent review with new analytical results for the NJL can be found in Ref. \cite {imc}. 
Another interesting result obtained in Ref. \cite {prd} concerns the size of the first order segment of the transition line which expands with increasing $B$ in such a way that the critical point becomes located at higher temperature and smaller chemical potential values. Note that, depending on the adopted parametrization, this region can display a rather complex pattern with multiple weak first order transitions taking place \cite {norberto}.

Concerning the low temperature portion of the phase diagram 
one notices that so far most applications have considered effective models with scalar and pseudoscalar channels only. However, as already pointed out, the presence of a vector interaction 
can be an important ingredient to reproduce some experimental results or compact star observations, and so, should also be taken into  
account in the computation of the EOS for magnetized quark matter.  
A step towards this type of investigation has been recently taken in Ref. \cite {robson}  where  two flavor magnetized quark matter in the presence of a repulsive vector coupling, described by the NJL model, has been considered. The results show that the vector interaction counterbalances the effects produced by a strong magnetic field. For instance, in the absence of the vector interaction,  high magnetic fields ($eB \ge 0.2 \, {\rm GeV}^2$) increase  the first order transition region. On the other hand  a decrease of this region is observed for a strong vector interaction and vanishing magnetic fields. Also, at low temperatures and $G_V=0$, the coexistence  chemical potential decreases with an increase of the magnetic field (IMC) \cite{prd}, however, the inclusion of a the vector interaction results in the opposite effect.  
The presence of a magnetic field together with a repulsive vector interaction gives rise to a peculiar transition pattern since $B$ favors the appearance of multiple solutions to the gap equation whereas the vector interaction turns some metastable solutions into stable ones allowing for a cascade of transitions to occur  \cite {robson}. The most important effects take place at intermediate and 
low temperatures affecting the location of the critical end point as well as the region of first order chiral transitions. 

More realistic physical applications require that one considers more sophisticated versions of the simple two flavor model considered in Ref. \cite {robson}. 
Strangeness is a necessary ingredient when describing the structure of  compact stellar objects or the QCD phase diagram. Therefore, the purpose of the present work is to study magnetized strange quark matter in the presence of a repulsive vector interaction. We are also interested in understanding the properties of  
strongly interacting matter described by two different vector interactions 
\cite{klimt,hanauske} and \cite{fuku08} and two commonly used parametrizations 
of the NJL model \cite{hatsuda,reh}. In the following we refer to the 
extended version of the NJL model that incorporates a vector interaction
as NJLv model. We first evaluate the similarities and 
differences at zero temperature of pure quark matter obtained with 
the two models by investigating the behavior of the constituent quark masses  
and the related EOS for two different physical situations, namely matter 
with the same quark chemical potentials and the same quark densities.
Once the underlying physics is understood, we move to stellar matter  
conditions.  Having in mind two recently $2M_\odot$ pulsars measured 
PSR J1614-2230 \cite{Demorest10}, $1.97\pm 0.04\, M_\odot$, and   
PSR J0348+0432 \cite{j0348}, $2.01\pm 0.04\, M_\odot$, we discuss which form 
of the vector interaction results in higher compact star masses. 
 We devote special attention to the zero temperature
 part  of the phase diagram which is currently not accessible to lattice
 simulations and which constitutes the important region as far as the physics
 of compact stars is concerned. We do not consider the color
   superconducting phase in the interior of hybrid stars, which would make the
   equation of state softer. Our conclusions on the maximum star masses  should, therefore, be regarded as upper limits.

 \section{General formalism}

In order to consider quark matter under the influence of  
strong magnetic fields and in the presence of a repulsive vector interaction we introduce the following Lagrangian density, 
where the quark sector is described by the  SU(3) version of the 
NJL model: 
$${\cal L} = {\bar{\psi}} \left[\gamma_\mu\left(i\partial^{\mu} 
- q A^{\mu} \right)- 
{\hat m}_f \right ] \psi ~+~ {\cal L}_{sym}~ \nonumber $$ 
\begin{equation} 
+~{\cal L}_{det}~+{\cal L}_{vec} - \frac {1}{4}F_{\mu \nu}F^{\mu \nu}, 
\label{njl} 
\end{equation} 
where ${\cal L}_{sym}$ and ${\cal L}_{det}$ are given by: 
\begin{equation} 
{\cal L}_{sym}~=~ G_S \sum_{a=0}^8 \left [({\bar \psi} \lambda_ a \psi)^2  
+ ({\bar \psi} i\gamma_5 \lambda_a \psi)^2 \right ]  ~, 
\label{lsym} 
\end{equation} 
\begin{equation} 
{\cal L}_{det}~=~-K \left \{ {\rm det} \left [ {\bar \psi}(1+\gamma_5) \psi  
\right] +  {\rm det} \left [ {\bar \psi}(1-\gamma_5) \psi \right] \right \} ~, 
\label{ldet} 
\end{equation} 
where $\psi = (u,d,s)^T$ represents a quark field with three 
flavors, ${\hat m}_f= {\rm diag} (m_u,m_d,m_s)$ is 
 the corresponding (current) mass matrix while $q$ 
represents the quark electric charge, $\lambda_0=\sqrt{2/3}I$  where 
$I$ is the unit matrix in the three flavor space, and 
$0<\lambda_a\le 8$ denote the Gell-Mann matrices. We consider 
$m_u=m_d \ne m_s$. The  ${\cal L}_{det}$ term is the t'Hooft 
interaction which represents a determinant in flavor space which, for 
three flavor, gives a six-point interaction \cite {buballa}  
and ${\cal L}_{sym}$, which is symmetric 
under global  $U(N_f)_L\times U(N_f)_R$  transformations and corresponds to a 
4-point interaction in flavor space. The parameters of the model,  
$\Lambda$, the coupling constants $G_S$ and $K$ 
and the current quark masses $m_u^0$ and $m_s^0$ are determined  by fitting 
$f_\pi$, $m_\pi$ , $m_K$ and $m_{\eta'}$ to their empirical values. 
Two parametrization sets are used in the present work and the constant 
values are given in Table \ref{tab1}. 
 
\begin{table}[h] 
\centering 
\begin{tabular}{cccccc} 
\hline   
Parameter set & $\Lambda$ & $G_S\Lambda^2$ & $K\Lambda^5$ & 
$m_{u,d}$ & $m_{s}$ \\ 
         &    MeV    &          &   &  MeV   &  MeV  \\ \hline 
HK \cite{hatsuda}& 631.4 & 1.835 & 9.29 & 5.5 & 135.7 \\ 
RKH \cite{reh} & 602.3 & 1.835 & 12.36 & 5.5 & 140.7 \\ 
\hline 
\end{tabular} 
\caption{\label{tab1} Parameter sets for the NJL SU(3) model.} 
\end{table} 
 
We employ a mean field approach and  
the effective quark masses can be obtained self consistently  from  
\begin{equation} 
 M_i=m_i - 4 G_S \phi_i + 2K \phi_j \phi_k,  
 \label{mas1} 
\end{equation} 
with $(i,j,k)$ being any permutation of $(u,d,s)$.  
 
As for the vector interaction, the Lagrangian density that 
denotes the $U(3)_V \otimes U(3)_A$ invariant interaction is 
\cite{ruggieri2009,bonanno,lenzi2012,odilon2012,lee2013}: 
\begin{equation} 
{\cal L}_{vec} = - G_V \sum_{a=0}^8  
\left[({\bar \psi} \gamma^\mu \lambda_a \psi)^2 + 
 ({\bar \psi} \gamma^\mu \gamma_5 \lambda_a \psi)^2 \right]. 
\label{p1} 
\end{equation} 
 
For the SU(2) version of the NJL model,  
at non-zero quark densities, the flavor singlet condensate term of the vector  
interaction, (${\bar \psi} \gamma^0 \lambda_0 \psi$),  
develops a non-zero expectation value while all other components of the vector  
and axial vector interactions have vanishing mean fields. Hence, a 
reduced NJLv Lagrangian density can be written as 
\cite{fukushima2008,weise2012,shao2013,sasaki2013,masuda2013}: 
 
\begin{equation} 
{\cal L}_{vec} = - G_V ({\bar \psi} \gamma^\mu \psi)^2. 
\label{p2} 
\end{equation} 
 
In the SU(3) NJLv model, the above Lagrangian densities are not identical  
in a mean field approach and we discuss both cases next. We refer to the 
Lagrangian density given in Eq. (\ref{p1}) as model 1 (P1) and to the
Lagrangian density given in Eq. (\ref{p2}) as model 2 (P2).

As  usual, $A_\mu$ and $F_{\mu \nu }=\partial 
_{\mu }A_{\nu }-\partial _{\nu }A_{\mu }$ are used to account 
for the external magnetic field. We are interested in a 
static and constant magnetic field in the $z$ direction and hence,  
we choose $A_\mu=\delta_{\mu 2} x_1 B$. 
 
We need to evaluate the thermodynamical potential for the three flavor 
quark sector, $\Omega_f$, which as usual can be written as  
$\Omega = -P = {\cal E} - T {\cal S} - \sum { \mu} \rho $ where $P$ 
represents the pressure, ${\cal E}$ the energy density, $T$ the temperature, 
${\cal S}$ the entropy density, and ${\mu}$ the chemical potential.  
To determine the EOS for the SU(3) NJL at finite density and in the presence 
of a magnetic field  we need to know the scalar condensates, $\phi_i$, 
the quark number densities, $\rho_i$, as well as 
the contribution from the gas of quasi-particles, $\theta_i$.  
In the presence of a magnetic field all these quantities have been evaluated  
with great detail in \cite{prc1,prc2}, from where the mathematical expressions
with vacuum, medium and magnetic field contributions can be obtained.

If model 1 is considered, the pressure reads: 
 
$$ P = \theta_u+\theta_d+\theta_s  
-2G_S(\phi_u^2+\phi_d^2+\phi_s^2) + \nonumber $$ 
\begin{equation} 
2G_V ( \rho_u^2 + \rho_d^2 +\rho_s^2) + 4K \phi_u \phi_d \phi_s \,\,, 
\label{pressp1}
\end{equation} 
and the effective chemical potential, for each flavor, is given by  
\begin{equation} 
{\tilde \mu}_i = \mu_i - 4 G_V \rho_i. \quad i=u,d,s 
\label{mup1}
\end{equation} 
We also refer to P1 as the flavor dependent model, for the reasons that
will become obvious from the analysis of our results. 

If, on the other hand,  model 2 is considered, the pressure becomes: 
 
$$ P = \theta_u+\theta_d+\theta_s  
-2G_S(\phi_u^2+\phi_d^2+\phi_s^2) + \nonumber $$ 
\begin{equation} 
G_V \rho^2 + 4K \phi_u \phi_d \phi_s \,\,,
\label{pressp2}
\end{equation} 
where 
\begin{equation} 
\rho=\rho_u+\rho_d+\rho_s,  \quad \rho_B=\rho/3, 
\end{equation} 
and in this case the effective chemical potential, for each flavor, 
is given by  
\begin{equation} 
{\tilde \mu}_i = \mu_i - 2 G_V \rho. 
\label{mup2}
\end{equation} 
We next refer to P2 as the flavor independent (or flavor blind) model. In both cases note that, as pointed out in Ref. \cite {buballa}, ${\tilde \mu}_i$ is a strictly rising function of $\mu_i$.

If stellar matter is to be considered, $\beta$-equilibrium and charge 
neutrality have to be imposed and a leptonic sector is then included. The 
Lagrangian density reads: 
\begin{equation} 
\mathcal{L}_l=\bar \psi_l\left[\gamma_\mu\left(i\partial^{\mu} - q_l A^{\mu} 
\right) -m_l\right]\psi_l \,\,, 
\label{lage} 
\end{equation} 
where $l=e,\mu$ and the leptonic contributions to the pressure, density and 
entropy density are also given in \cite{prc1,prc2}.

\section{Results and discussions} 
 
We next analyze two different physical situations: pure quark matter,
of interest in the studies of the QCD phase diagram, and stellar matter
applied to investigate possible quark and hybrid stars.

\subsection{Pure quark matter} 
In the present section we discuss two distinct physical 
situations: quark matter defined by equal
chemical potentials for three flavors $u,\, d, \, s$
and for equal quark densities. 
We discuss the effect of the vector interaction on the EOS and
strangeness fraction. In particular, we take $G_V=x G_S$, where $x$ is a
free parameter which we vary such that $0 < x < 1$, as proposed
in \cite{hanauske}. We present results for both
possible forms of the vector interaction discussed in the previous
section, which are designated by P1 and P2, respectively, the
flavor dependent/independent form. We also compare two popular
parametrizations of the SU(3) NJL model designated by HK \cite{hatsuda}
and RKH \cite{reh}.

\begin{figure}[ht]
\begin{tabular}{cc}
\includegraphics[width=1\linewidth,angle=0]{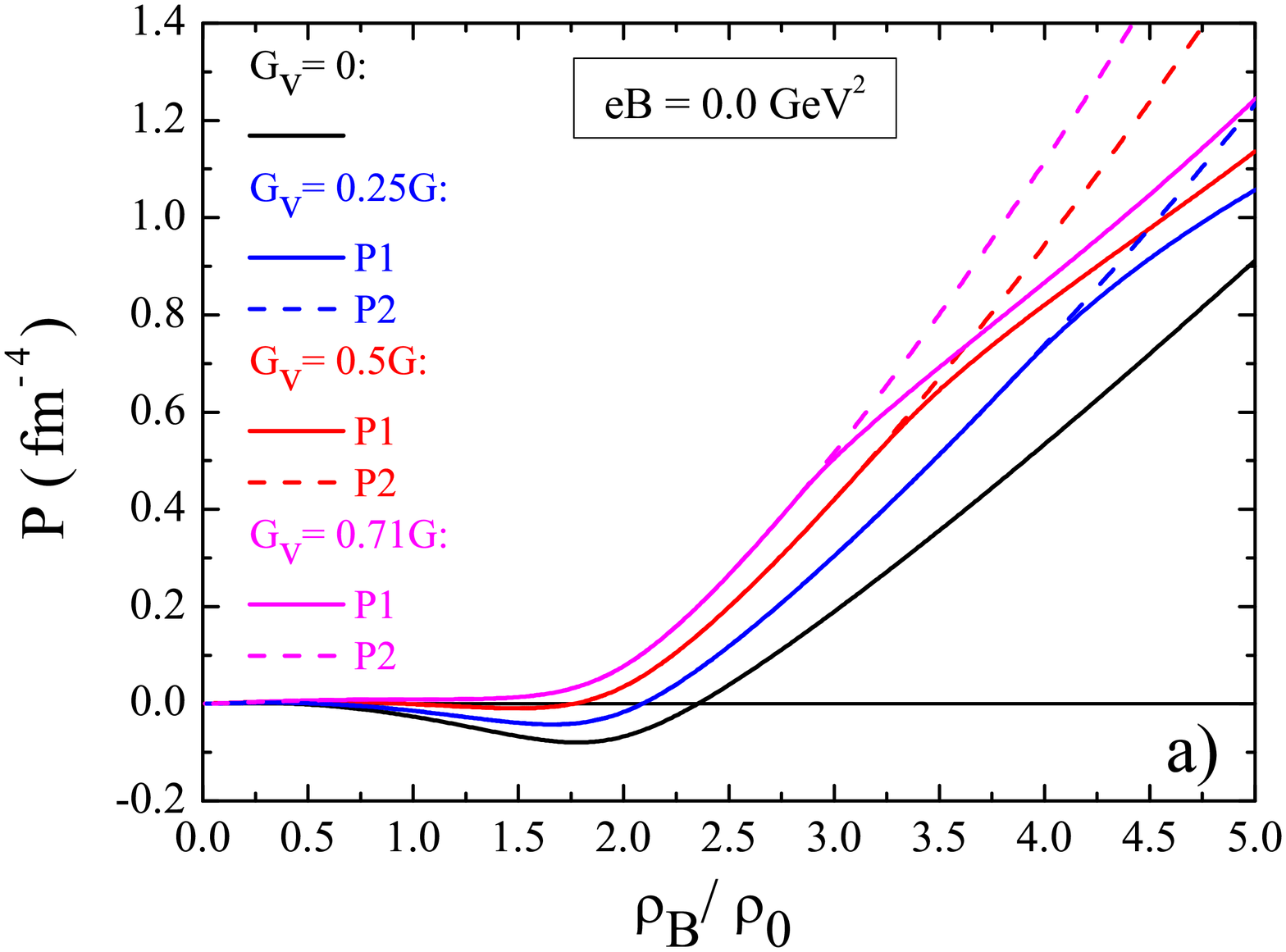}\\
\includegraphics[width=1\linewidth,angle=0]{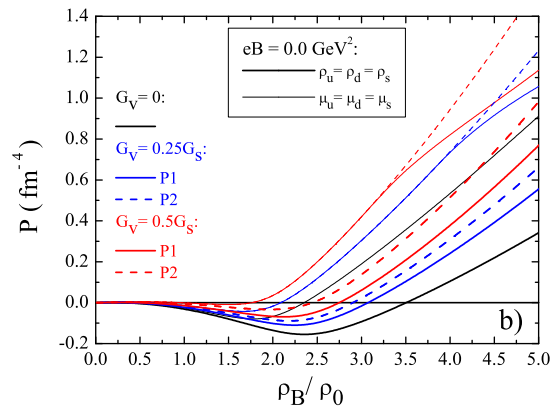}
\end{tabular}
\caption{The pressure versus baryonic density for model 1 (P1)
  and 2 (P2) for different values of $x$ and parametrization RKH, 
under the conditions a) $\mu_u=\mu_d=\mu_s$, 
b) $\rho_u=\rho_d=\rho_s$ (thick lines) and
  $\mu_u=\mu_d=\mu_s$ (thin lines).} 
\label{fig1}
\end{figure}

The effect on the EOS of the different forms for the vector
interaction is seen in Fig. \ref{fig1}a), where the parametrization RKH
is used with different strengths of the vector interaction, for both
P1 and P2 under the equal chemical potentials constraint. Several
conclusions are in order: a) the models coincide until $\sim
3-4\rho_0$, where $\rho_0=0.17$ fm$^{-3}$ is the nuclear matter saturation 
density, depending on the magnitude of $x$. The larger $x$ the
earlier the two models differ. This is due to the onset of the
strangeness that occurs at smaller densities with form P1 as 
is shown latter; b) once
the strangeness sets on the EOS becomes softer, therefore, for large
enough densities P1 is softer than P2; c) the pressure is negative 
for some values of $G_V$, including $G_V=0$, a feature observed and
discussed in \cite{hanauske}, with consequences on possible 
coexisting phases and associated phase transition; d) for a
large enough $G_V$ the first order phase transition observed for
densities below 2$\rho_0$ disappears, and the pressure increases
monotonically with the baryonic density. For the parametrization RKH this
occurs for $x=0.71$ and is represented by the pink curves in the
figure.

In  Fig. \ref{fig1}b), we compare two different scenarios,
$\mu_u=\mu_d=\mu_s$ and  $\rho_u=\rho_d=\rho_s$ represented,
respectively, by the thin and thick lines. The equal flavor densities,
corresponding to matter generally designated by strange quark matter,
is softer, gives rise to a larger density discontinuity at the first
order phase transition. In this scenario the EOS for models P1 and
P2 differ for all baryonic densities because the vector interaction
form given in Eq. (\ref{p2}) results in different contributions in each case.
This scenario may be approximately realized at the center of a quark star. 

\begin{figure*}[ht]
\begin{tabular}{cc}
\includegraphics[width=0.75\linewidth,angle=0]{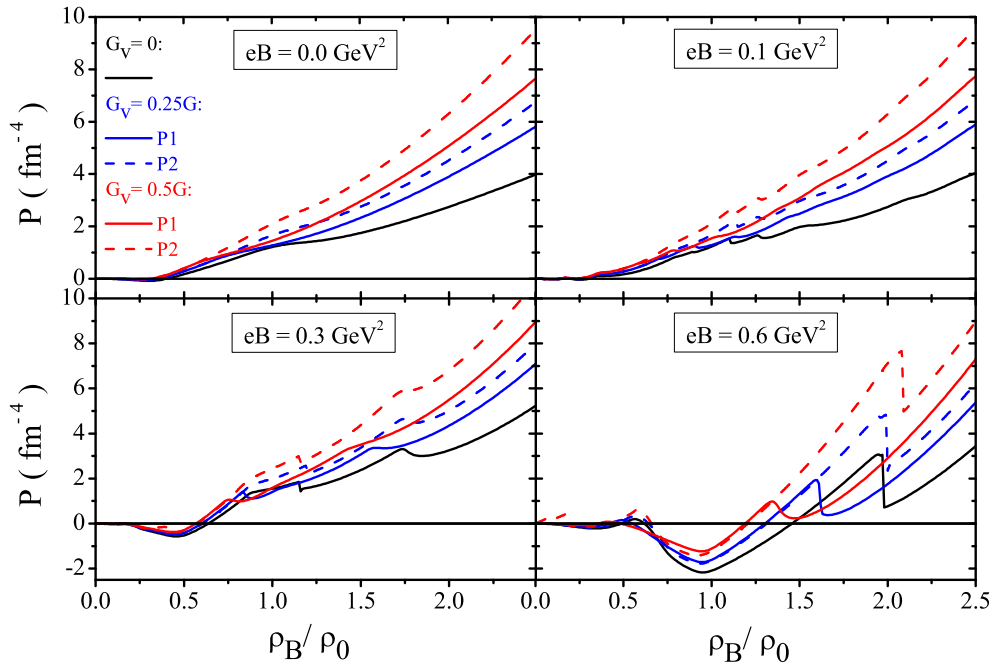}\\
\end{tabular}
\caption{Pressure versus baryonic density for equal chemical
  potentials and models P1 and
  P2 for different values of x, and several intensities of the
  magnetic field: $eB=0$, $eB=0.1$ GeV$^2$, $eB=0.3$ GeV$^2$ and $eB=0.6$ GeV$^2$.}
\label{fig2}
\end{figure*}

The effect of the magnetic field on the EOS is seen comparing the four
graphs of Fig. \ref{fig2}. We first discuss the scenario
$\mu_u=\mu_d=\mu_s$.  We have chosen  three values of $eB$, 0.1,
0.3 and 0.6 GeV$^2$ corresponding to $5m_\pi^2$,  $15m_\pi^2$ and  $30
m_\pi^2$. The van-Alphen oscillations due to the filling of the Landau
levels are already seen for $eB=0.1$ GeV$^2$.  The EOS becomes harder
at large densities, and the larger $eB$ the harder the EOS, although
locally, when the filling of a new Landau level begins, the EOS
becomes softer. This increased softness is immediately overtaken by an
extra hardness. The larger $B$ the larger the amplitude of the
fluctuations and the smaller the number of them, because less Landau
levels are involved. The softening occurring when a new Landau level
starts being occupied has a strong effect at the smaller densities
giving rise to a pressure that is negative within a larger range of
densities. For $eB=0.3$ GeV$^2$, a magnetic field that could occur 
at LHC experiments, negative pressures occur beyond 
$\rho_B=0.5$ fm$^{-3}$ and this range increases until $\sim 1-1.5$fm$^{-3}$ 
for $eB=0.6$ GeV$^2$. The vector interaction
P2 gives always the hardest EOS due to the smaller strangeness content.

\begin{figure}[th]
\begin{tabular}{cc}
\includegraphics[width=0.8\linewidth,angle=0]{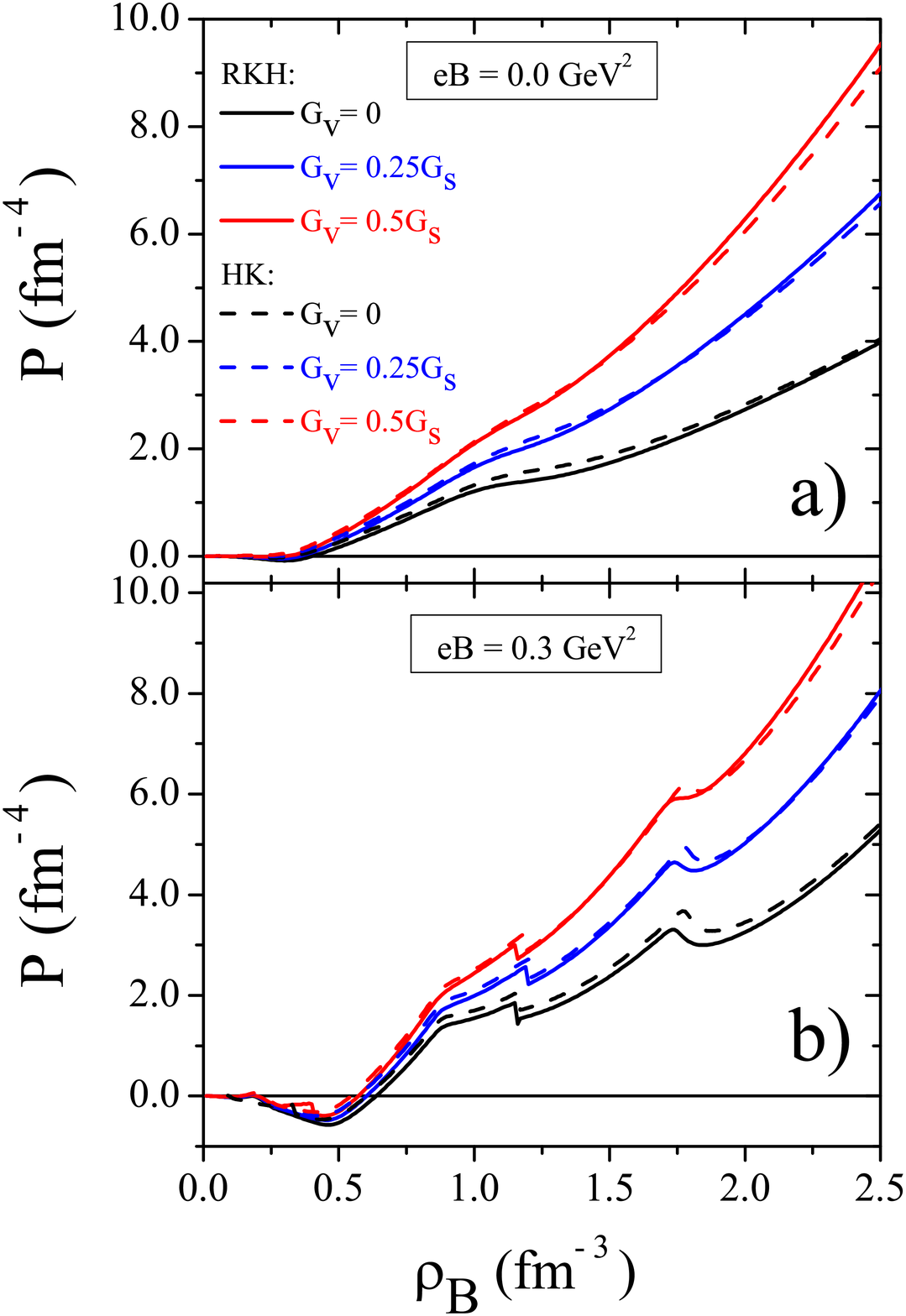}
\end{tabular}
\caption{Pressure versus baryonic density for  model
  2 (P2) for different values of $x$. Two parametrizations of the NJL
  are compared HK and RKH with magnetic field intensities:
  a) $eB=0$; b) $eB=0.3$ GeV$^2$.}
\label{fig3}
\end{figure}

In Fig. \ref{fig3} the EOS obtained with interaction P2 and the two
different parametrizations of the NJL model are compared for $eB$=0
and 0.3 GeV$^2$. For $G_V=0$ the EOS obtained with the HK parametrization 
does not cross the RKH EOS.
This is no longer valid for a finite $G_V$. The  RKH EOS becomes
stiffer and the two EOS cross within the range of densities shown in the 
figure. This feature is still
present for a finite magnetic field (see Fig. \ref{fig3}b). 

\begin{figure}[ht]
\begin{tabular}{cc}
\includegraphics[width=0.8\linewidth,angle=0]{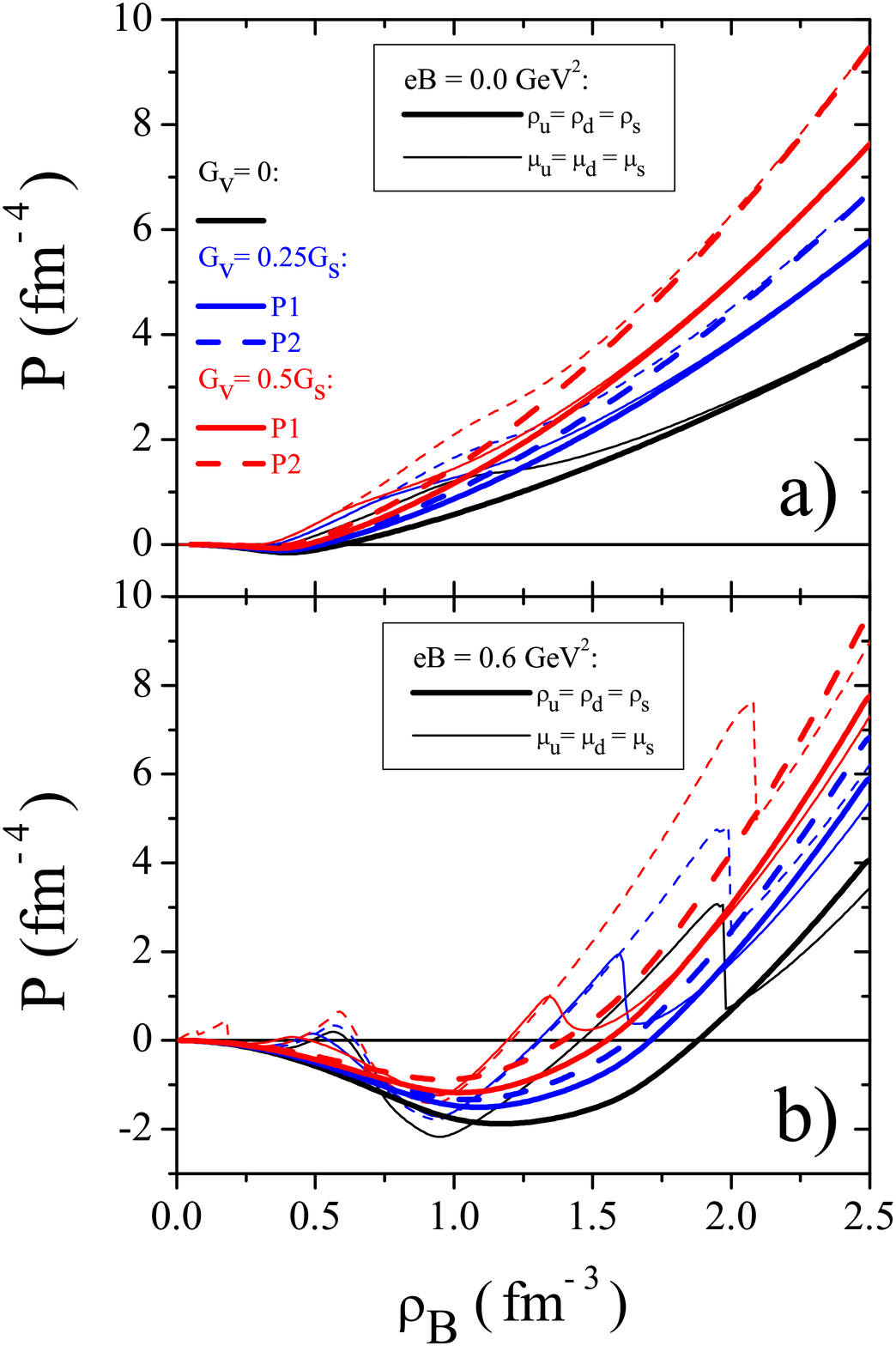}
\end{tabular}
\caption{Pressure versus baryonic density for models P1 and P2 for equal 
quark  densities,  different values of $x$ for: a) $eB=0$; b) $eB=0.6$ GeV$^2$.}
\label{fig4}
\end{figure}

Now we move to the scenario of equal flavor densities. The EOS are 
plotted in Fig. \ref{fig4}
for $eB=0$ and $0.6$ GeV$^2$. As already referred before,  this scenario
is softer than the equal chemical potentials one for the range of
densities shown. However, at sufficiently large densities both
scenarios converge. In fact, above chiral symmetry restoration it is
expected that equal chemical potentials correspond to equal
densities. The effect of a strong magnetic field is very different in
both scenarios: while the equal chemical potentials EOS presents very
strong oscillations, these are not seen for the scenario of equal
densities. In the equal chemical potentials the $s$-quark density
remains zero until a quite high baryonic density, and, therefore, for
a given density below the strangeness onset the $u$ and $d$-quark
densities are much larger than in the equal quark densities. 
Larger $u$ and $d$ quark densities give rise to the 
restoration of chiral symmetry at lower baryonic densities. Since the
effect of the magnetic field is stronger the smaller the masses, this
explains the differences in the bottom graphs of Fig. \ref{fig4} 
between the two scenarios.

\begin{figure*}[ht]
\begin{tabular}{cc}
\includegraphics[width=0.75\linewidth,angle=0]{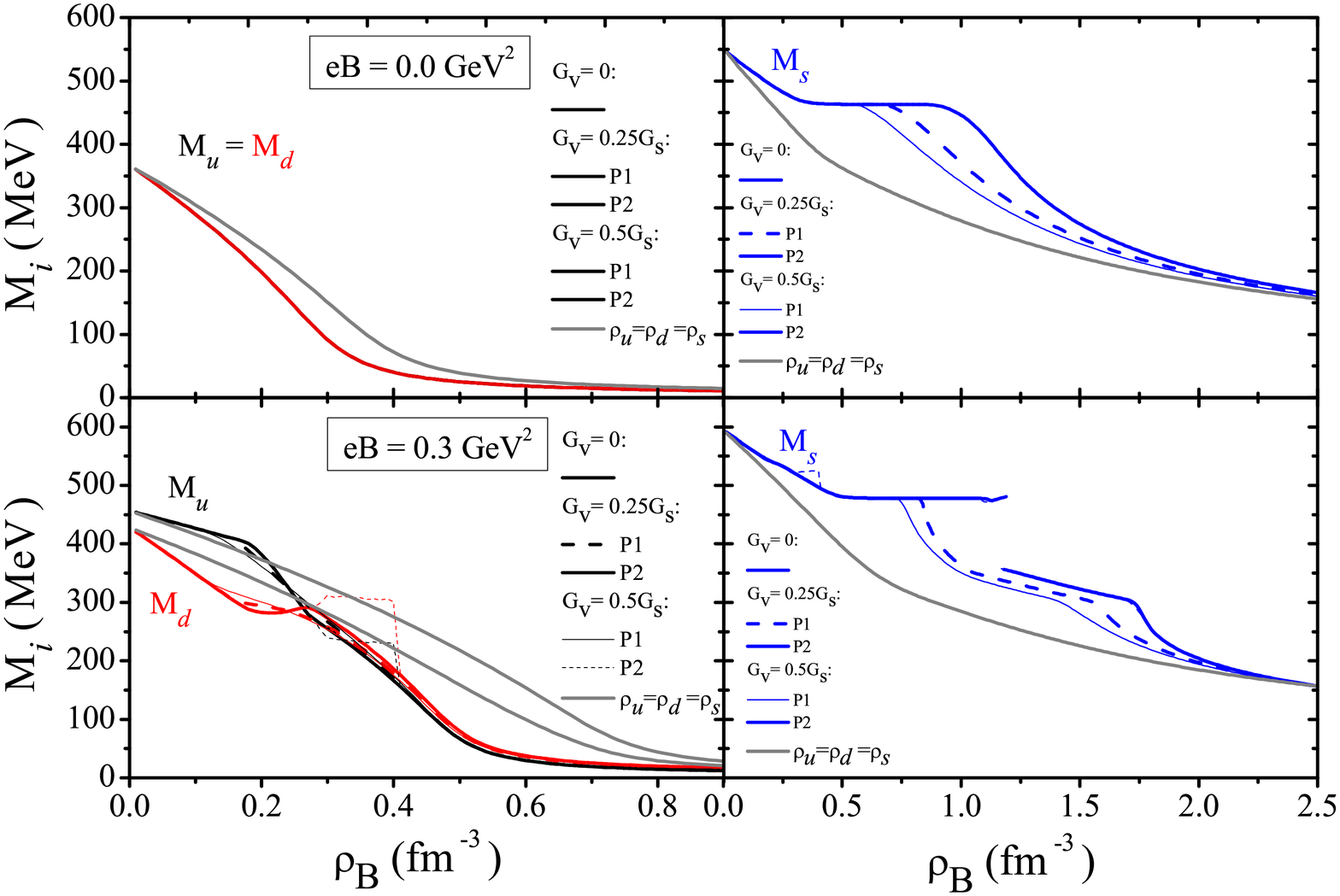}
\end{tabular}
\caption{The quark constituent masses as a function of the baryonic
  density  for models P1 and P2, different values of $x$ for 
  $eB=0$ (top figures) and $eB=0.3$ GeV$^2$ (bottom figures).}
\label{fig5}
\end{figure*}

The difference between the chiral symmetry restoration in the
two scenarios presented above is clearly seen in Fig. \ref{fig5},
where the constituent masses of the $u$, $d$, and $s$ quarks are
plotted for different strengths of the vector interaction and the two
models P1 and P2. We first comment on the $eB=0$ results and the two
vector interactions, top panels of Fig. \ref{fig5}. The chiral restoration
of $u$ and $d$ quarks does not depend on the interaction. However, a 
difference is observed between the equal chemical potentials and equal 
densities scenarios.

In the scenario of equal densities (gray lines), one can see that the
chiral symmetry restoration of the $u$ and $d$ quarks occurs at 
larger densities than in the situation with equal chemical potentials
(red lines) because the $u$ and $d$ quark densities are larger in the last situation. For the
$s$ quark, the opposite occurs. Including the vector interaction does not
  affect the quark masses in model P2, but it does affect the $s$ quark mass
  in model P1. In this case the larger $G_V$ the faster the chiral restoration
  of the $s$-quark mass, due to the larger $s$-quark density.
At finite $B$ similar conclusions are drawn, but also new aspects arise.
First of all the constituent masses of $u$ and $d$ quarks do not
coincide anymore due to the charge difference.  Since the $u$ quark
has a larger charge, $M_u>M_d$ in  the
scenario of equal densities. In the scenario of equal chemical
potentials there is a competition between the effect of the charge and
the effect of density. For the larger magnetic field considered
discontinuities are obtained. These correspond to first order phase
transitions associated to the filling of the Landau levels.

 The above results on the constituent quark masses confirm that the
large oscillations of the EOS seen in Fig. \ref{fig4}b)  for the equal
chemical potentials is in fact due to the small masses of the $u$ and
$d$ quarks.

\begin{figure}[ht]
\begin{tabular}{cc}
\includegraphics[width=0.8\linewidth,angle=0]{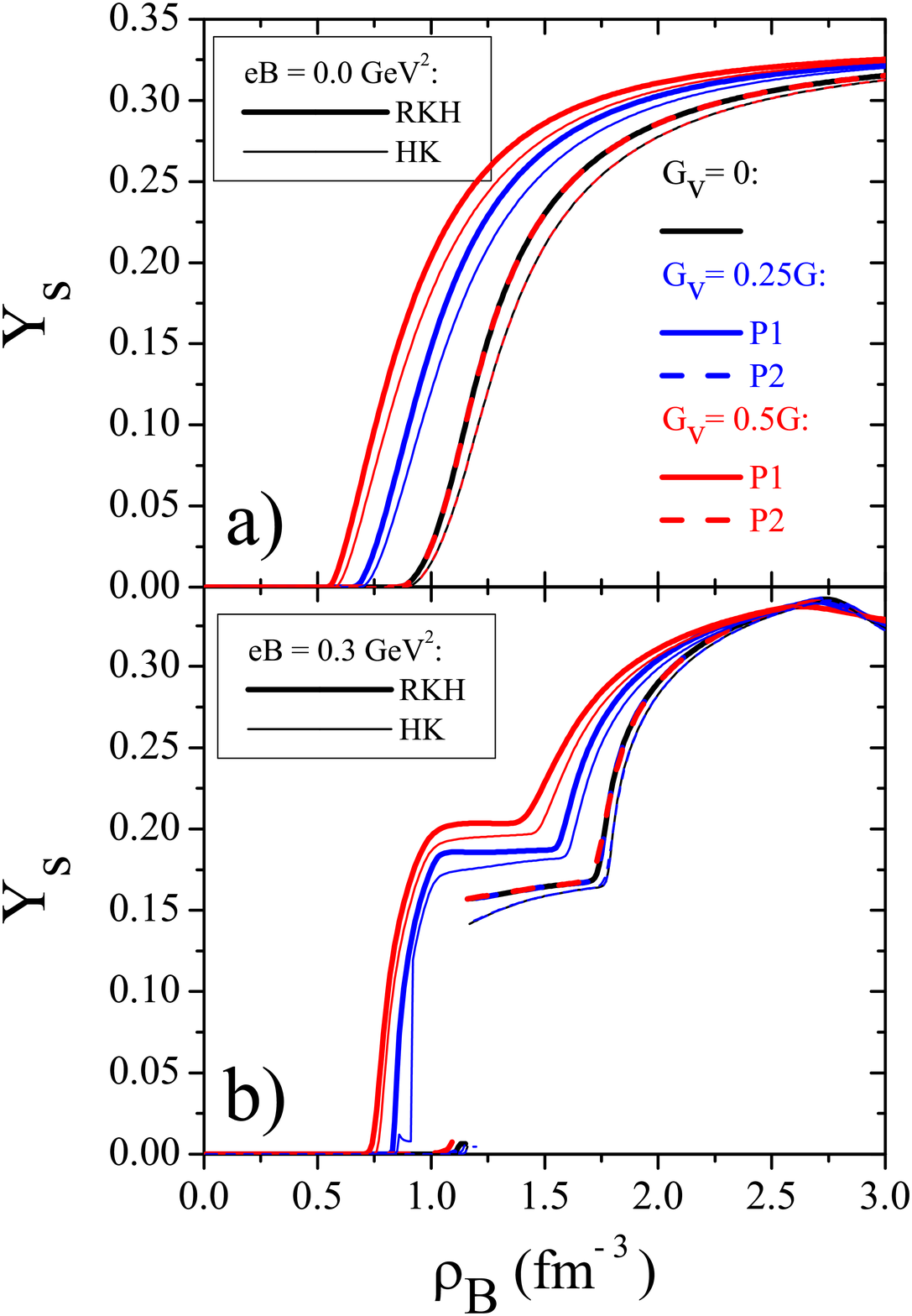}
\end{tabular}
\caption{The strangeness fraction as a function of the baryonic
  density  for models P1 and P2 and different values of $G_V$, and a) $B=0$; b)
 $eB=0.3$ GeV$^2$.
}
\label{fig6}
\end{figure}

It is interesting to compare the strangeness content of matter under
the conditions discussed until this point.
We next analyze once more the situation of equal chemical potentials.
In Fig. \ref{fig6} the strange quark fraction for P1 and P2 models,  three
values of $G_V$, and   $eB=0$ and 0.3 GeV$^2$
are displayed. We also report
results for the parametrizations RKH (thick) and HK (thin). One aspect that
is immediately observed is that the P2 model presents the least amount of
strange quarks,  and its content does not depend on $G_V$, both for
zero  and a finite  magnetic field. However, the P1 model does
affect the strange quark content and the larger $G_V$ the earlier is
the onset of the $s$-quark, and the larger its content. 
One should notice that the definition of the effective chemical 
potentials in equations (\ref{mup1}) and (\ref{mup2}) is directly reflected 
on the strangeness content.
The magnetic field does not erase this feature.  Nevertheless, the filling
of new Landau levels decreases the rate of the increase of the $s$-quark
content as observed in Fig. \ref{fig6}b). Parametrizations RKH and HK
behave in a similar way with RKH predicting an onset of $s$-quarks at
smaller densities, and a larger amount of strangeness for a given
baryonic density.

\subsection{Stellar matter: quark stars}

We next move to the study of stellar matter, i.e., matter where
$\beta$-equilibrium and charge neutrality are enforced. In this case, 
leptons are introduced in the system, so that equations
\begin{equation}
\mu_s=\mu_d=\mu_u+\mu_e, \qquad  \mu_e=\mu_\mu.
\label{qch}
\end{equation}
and
\begin{equation}
\rho_e+\rho_\mu=\frac{1}{3}(2\rho_u-\rho_d-\rho_s).
\label{chneutrality}
\end{equation}
are satisfied.
\begin{figure}[t]
\includegraphics[width=0.8\linewidth,angle=0]{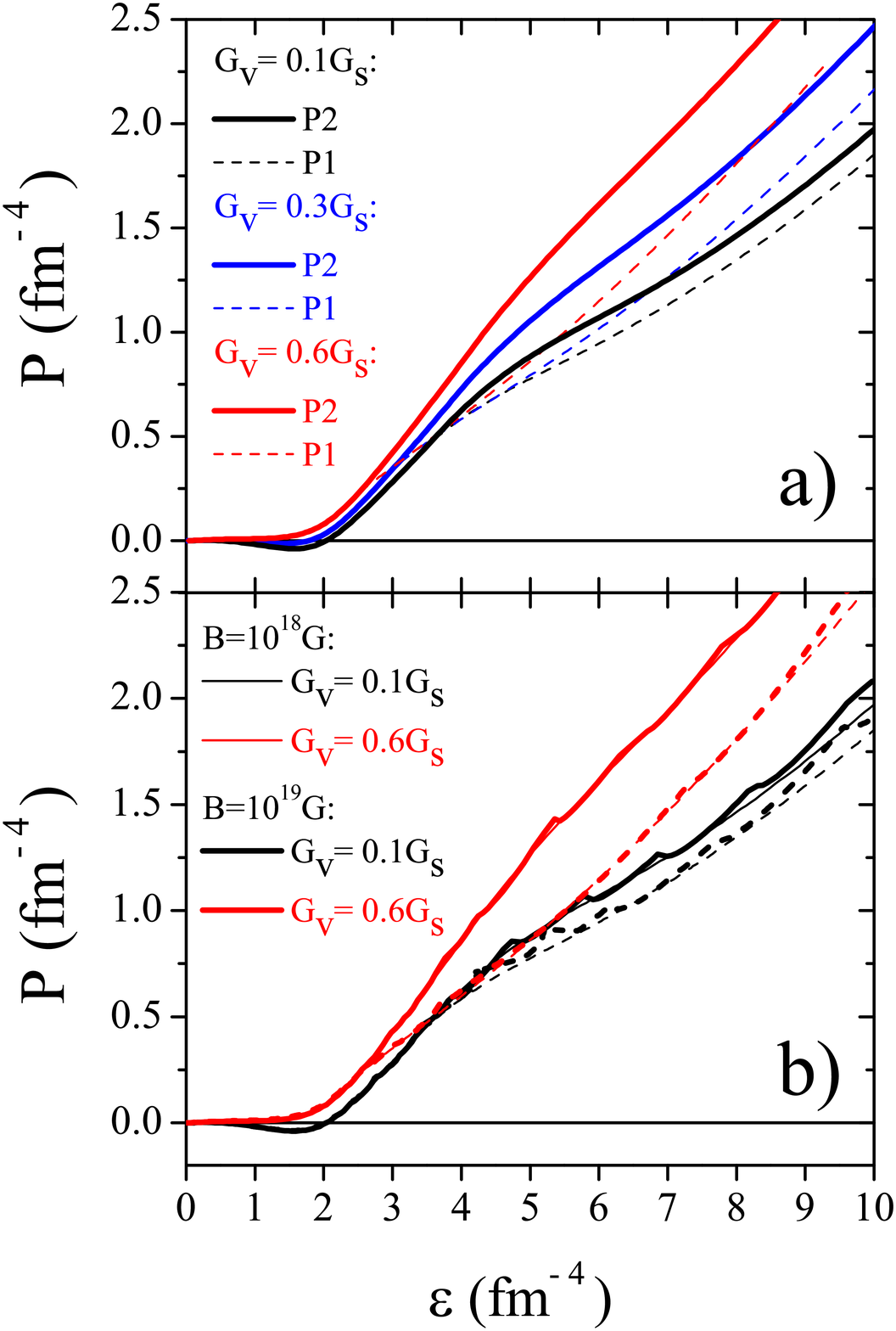}
\caption{The pressure versus energy density (EOS) for model P1 (thin lines)
  and P2  (thick lines)  for different values of $G_V$ and a) $B=0$; b)
  $B=10^{18}$ and  $10^{19}$ G. Both figures were obtained for the
parametrization RKH.}
\label{figs1}
\end{figure}

Since, to date, there is no information available on the star interior magnetic 
field, we assume that the magnetic field is baryon density-dependent as 
suggested in \cite{chakra97}.
In the following we consider a magnetic field that increases with density
according to 
\begin{equation}
B=B_{surf} +B_0(1-\exp[-\beta {(\rho/\rho_0)}^\gamma]), \quad
\beta=0.02,\, \gamma=3,
\label{mag}
\end{equation}
$B_{surf}=10^{15}$ G is the magnetic field at the surface of the star.
As our aim in this section is to compare
results with astrophysical observations, the use of magnetic fields in Gauss 
units is more adequate. We have considered that  $eB=1$ GeV$^2$ corresponds to 
$B=1.685 \times 10^{20}$ G. In the following we start by investigating the
effects of the vector interaction in stellar matter applied do quark stars and
subsequently we choose the best possible model and parameter set to build
hybrid stars and look at their macroscopic properties.

\begin{figure}[ht]
\begin{tabular}{cc}
\includegraphics[width=0.8\linewidth,angle=0]{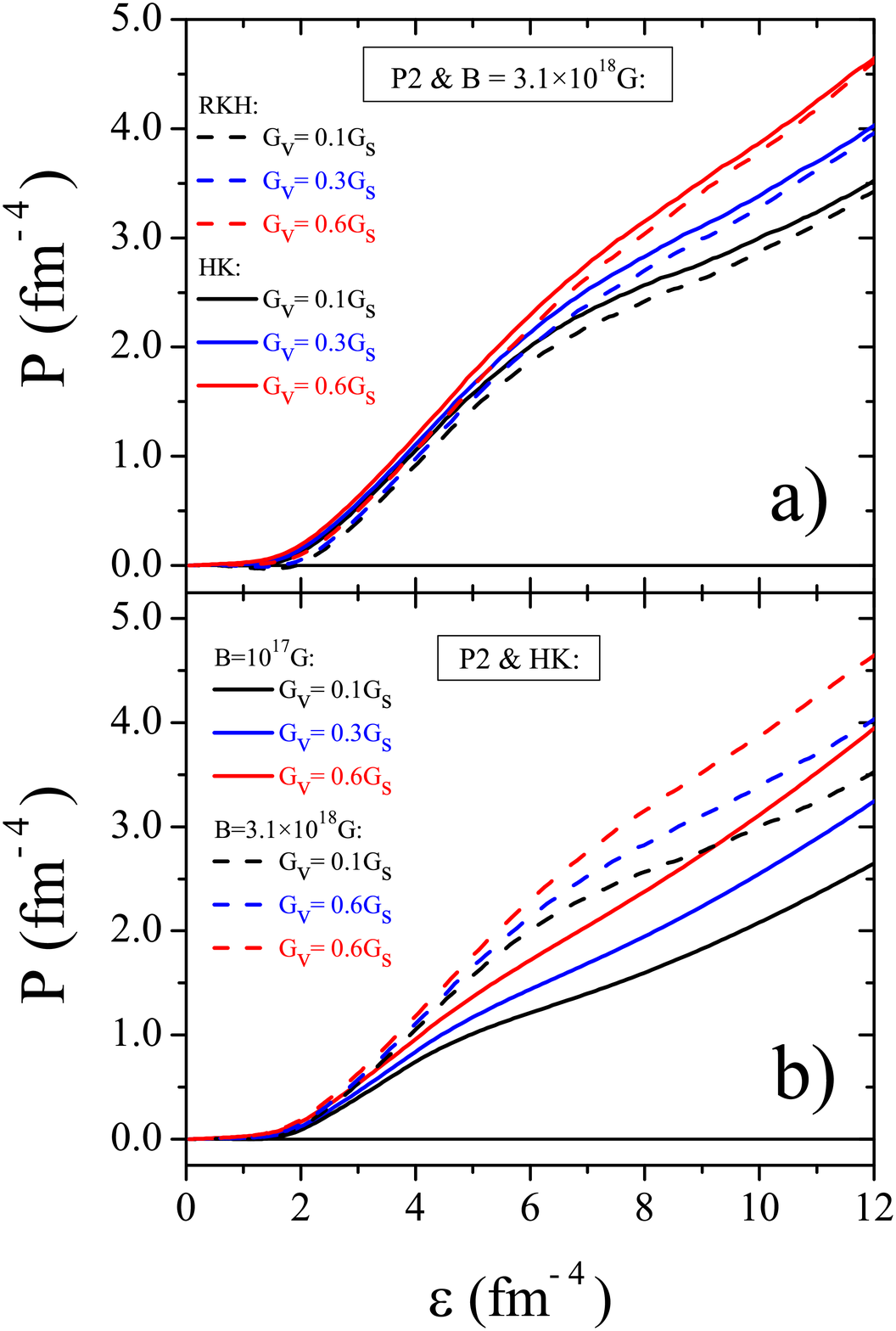}
\end{tabular}
\caption{EOS for model 2 (P2) for different values of $G_V$, 
and a) $B=3.1 \times B=10^{18}$ G obtained with 
parametrizations HK and RKH and 
b) two intensities of the magnetic field:
$B=10^{17}$ G and $B=3.1 \times 10^{18}$ G for parameter set HK.}
\label{figs3}
\end{figure}

Once again, we start from the non-magnetized case and check the differences
arising from both models with the RKH parameter set and different values of 
$G_V$ in Fig. \ref{figs1}. The same conclusions reached from the pure quark 
matter case can be drawn here, mainly that P2 gives rise to a harder EOS and 
that at very low energy densities, the pressure becomes slightly negative.
This difference can be easily understood if one looks at Eqs. 
(\ref{pressp1}) and (\ref{pressp2}), from where it is seen that the 
contribution from the vector term to the pressure is larger in model
P2 because in this case it is flavor blind. The effect of the
magnetic field on the quark matter is stronger for the large densities
when the magnetic field is more intense due to the density dependence
we have considered, see Eq. (\ref{mag}), and much larger when we
consider $B_0=10^{19}$ G.  
The fluctuations arising due to the filling of new Landau levels seem
larger and more frequent for the smaller vector coupling on an energy density
versus pressure curve. This arises because for a stronger vector term
a larger energy density is obtained for the same density, and
therefore, the fluctuations are spread over a larger energy density range. 

We then reobtain the EOS for the cases where $B=10^{17}$ and 
$3.1 \times 10^{18}$ G.  These values were chosen as the limiting ones 
because below $B=10^{17}$ G, all EOS coincide with the non-magnetized case and
 $3.1 \times 10^{18}$ G is the maximum value that allows us to avoid 
anisotropic pressures \cite{veronica}.
This is also the maximum intensity supported by a star bound by
  the gravitational interaction before the star becomes unstable \cite{lai91}.
 However, as we are using a density 
dependent magnetic field, this value may be never reached in the star core.

In Fig.\ref{figs3}a), we compare both parametrizations for a fixed magnetic
field equal to $3.1 \times 10^{18}$ G and different values of $G_V$. 
We can observe that HK yields harder 
EOS than RKH. The van Alphen oscillations are noticeable for this field 
intensity. The feature of HK and RKH EOS crossing with the increase
of the vector interaction, observed when pure quark 
matter is analyzed, occurs at energy densities larger than the ones shown in 
the figure.
In Fig. \ref{figs3}b), we fix the HK parametrization and plot the EOS for 
the two intensities of the magnetic field mentioned above. It is interesting to
observe that at large densities an EOS obtained with a smaller magnetic field
becomes harder for certain values of $G_V$ than an EOS obtained with a much 
stronger magnetic field and a smaller value of $G_V$.  

\begin{figure}[ht]
\begin{tabular}{cc}
\includegraphics[width=0.8\linewidth,angle=0]{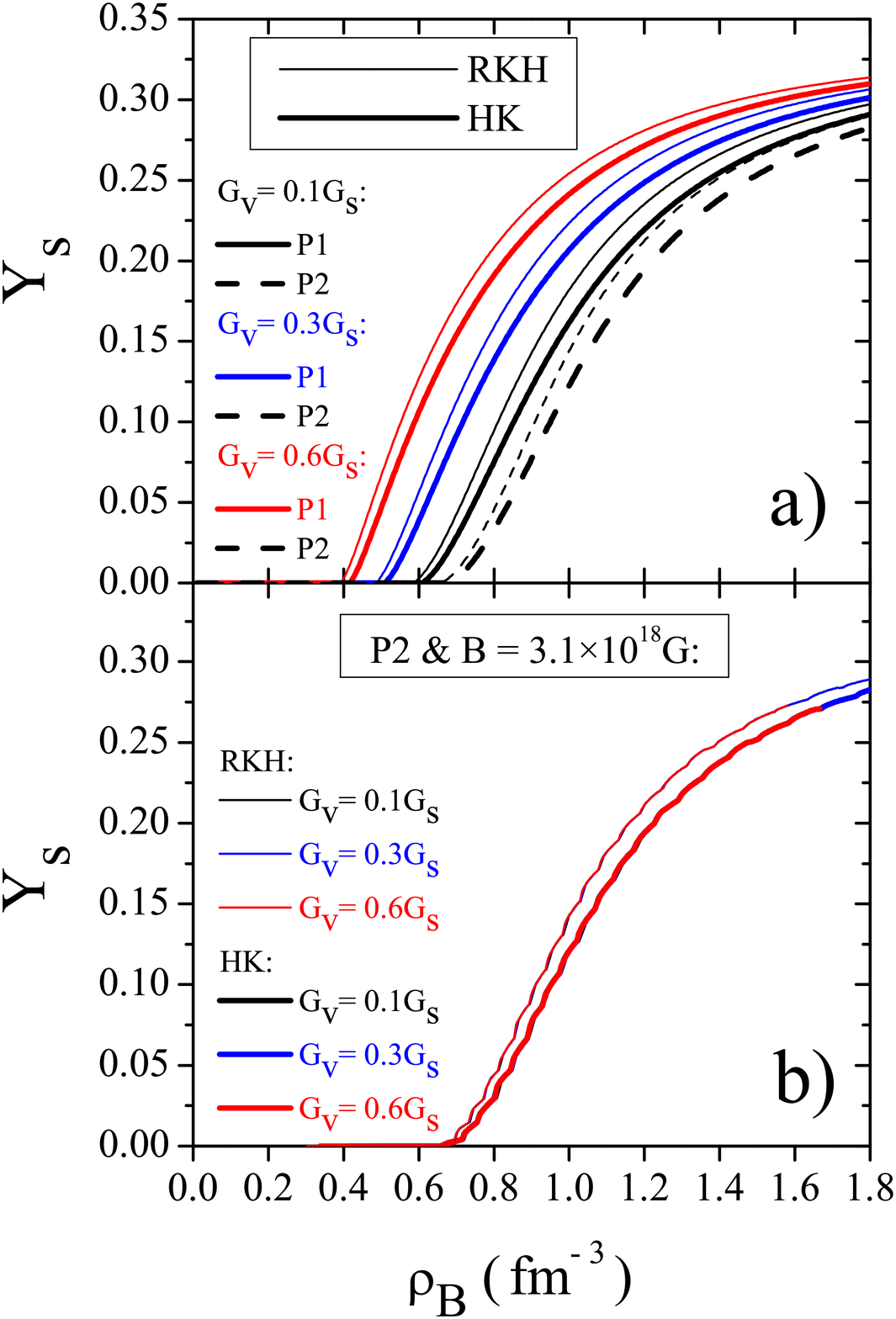}
\end{tabular}
\caption{Strangeness fraction as a function of the baryonic
density for $B=0$, parameter sets HK and RKH
a) for models 1 (P1) and 2 (P2) and different values of $x$ and
b) model 2 (P2) for different values of $x$ with $B=3.1 \times 10^{18}$ G.}
\label{figs2}
\end{figure}

\begin{table*}
\begin{tabular}{|c|c|c|c|c|c|c|c|c|}
\hline
 \multicolumn{3}{ |c| }{} &\multicolumn{3}{ |c| }{HK}&\multicolumn{3}{ |c| }{RKH} \\
  \hline
 \multicolumn{3}{ |c| }{} & $x=0.1$ & $x=0.3$ & $x=0.6$ & $x=0.1$ & $x=0.3$ & $x=0.6$ \\
 \hline
$B=0$~G  & P1 &  $M_{max}$ ($M_{0}$) & 1.49 & 1.58 & 1.69 & 1.27 & 1.35 & 1.46   \\
\cline{3-9}
 &   & R (km) & 9.13 & 10.89 & 11.98 & 8.01 & 8.17 & 9.41 \\
\cline{3-9}
 &   & $\varepsilon_{c}$ ($\mathrm{fm}^{-4}$) & 7.23 & 6.96 & 6.52 & 9.42 & 9.61 & 9.84 \\
\cline{3-9}
\cline{2-9}
 & P2 &  $M_{max}$ ($M_{0}$) & 1.56 & 1.72 & 1.91 & 1.35 & 1.54 & 1.74   \\
\cline{3-9}
 &  & R (km) & 9.15 & 10.61 & 11.47 & 8.22 & 8.60 & 9.91 \\
\cline{3-9}
 &  & $\varepsilon_{c}$ ($\mathrm{fm}^{-4}$) & 7.35 & 7.37 & 6.92 & 8.71 & 8.58 & 8.09 \\
\cline{3-9}
\hline
\hline
\multicolumn{2}{ |c| }{$B=10^{17}$~G} & $M_{max}$ ($M_{0}$) & 1.56 & 1.72 & 1.91 & 1.35 & 1.54 & 1.74 \\
\cline{3-9}
\multicolumn{2}{ |c| }{P2} & R (km) & 9.16 & 10.16 & 10.95 & 8.21 & 8.58 & 9.60 \\
\cline{3-9}
\multicolumn{2}{ |c| }{} & $\varepsilon_{c}$ ($\mathrm{fm}^{-4}$) & 7.41 & 7.36 & 6.98 & 8.80 & 8.94 & 8.11 \\
\cline{3-9}
\hline
\hline
\multicolumn{2}{ |c| }{$B=3.1\times10^{18}$~G} & $M_{max}$ ($M_{0}$) & 1.96 & 2.03 & 2.12 & 1.81 & 1.88 & 1.98 \\
\cline{3-9}
\multicolumn{2}{ |c| }{P2} & R (km) & 9.98 & 10.43 & 11.05 & 9.03 & 9.21 & 9.90 \\
\cline{3-9}
\multicolumn{2}{ |c| }{} & $\varepsilon_{c}$ ($\mathrm{fm}^{-4}$) & 7.41 & 7.22 & 6.78 & 8.74 & 8.21 & 7.80 \\
\cline{3-9}
\hline
\end{tabular}
\caption{Stellar macroscopic properties obtained from EOS of 
non-magnetized matter for models P1 and P2 and for magnetized matter with model
P2 and two values of magnetic field intensities. $M_{max}$ is the maximum mass,
R is the star radius and $\varepsilon$ the star central energy density.}
\label{table1}
\end{table*}

We proceed to the analysis of the strangeness content for
non-magnetized matter, whose curves are 
depicted in Fig. \ref{figs2}a). As in the case of pure quark 
matter, the amount of strange quarks remains unchanged with any variation of 
$G_V$ with model P2 while it increases with the increase of $G_V$ if model 
P1 is used. RKH presents higher strangeness content than HK with 
consequences in the maximum stellar masses, as we show next.
For the sake of completeness, we show the strangeness fraction for 
$B=3.1 \times 10^{18}$ G and the two parameter sets discussed in the 
present work in Fig. \ref{figs2}b) for the strangeness blind vector
interaction P2. As already expected from the softness of the EOS, we see that
HK introduces a smaller strangeness content in the system
and if we compare the values obtained with different values of the 
magnetic field ranging from $B=10^{17}$ G to $B=3.1 \times 10^{18}$ G, we 
can see that the amount of strange quarks remains practically unaltered for
both parameter sets.

\begin{figure}[ht]
\begin{tabular}{cc}
\includegraphics[width=0.8\linewidth,angle=0]{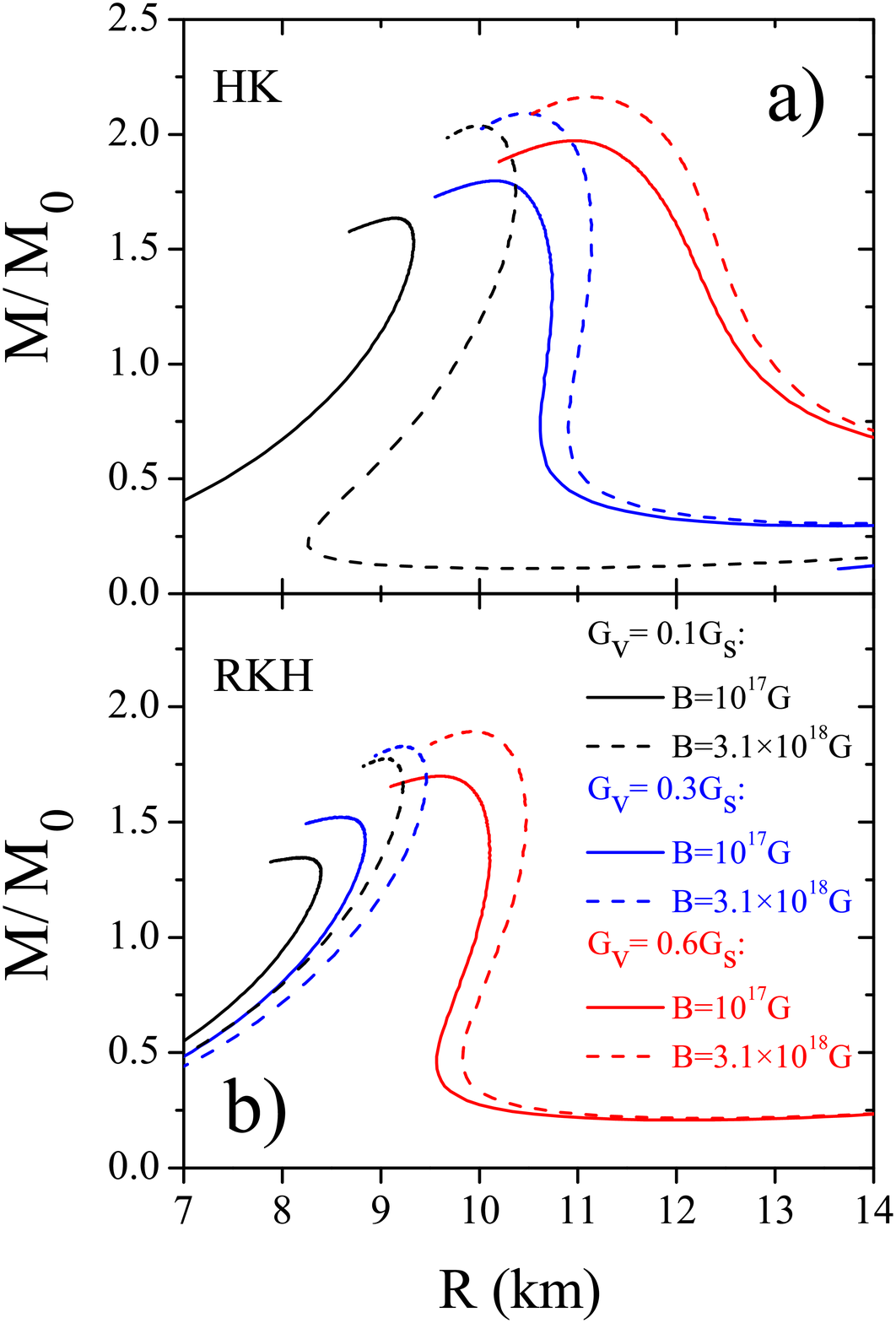}
\end{tabular}
\caption{Mass radius curves obtained with model 2 (P2) for different 
values of $G_V$, two intensities of the magnetic field ($B=10^{17}$ G and $B=3.1 \times 10^{18}$ G) and
parametrizations a) HK and b) RKH.}
\label{figs4}
\end{figure}

Finally, we use the EOS discussed above as input to the 
Tolman-Oppenheimer-Volkoff equations \cite{tov} and show our results in Fig. 
\ref{figs4} and Table \ref{table1}. A general trend is that HK, being harder 
with less strange quarks, produces higher maximum masses. A not so common 
feature is that for some combination of $G_V$ values and magnetic field 
intensities, the quark stars behave as hadronic stars in the sense that the
densities attained at low pressure are indeed very small. This is seen
in 
Fig.~\ref{figs4} in all cases where the low mass stars have very
large radii. This feature has already been observed in Ref.\cite{hanauske} 
for non-magnetized stars and it is related to the existence/non existence of 
negative pressures 
at very low densities for small/large values of the vector interaction
coupling.

We see that the maximum masses obtained with zero and 
low magnetic field intensities ($B=10^{17}$ G) are always coincident, but
the radii are slightly different due to the small differences in the central
energy densities.
Within  RKH the most  massive neutron stars have less $\sim 0.2\,
M_\odot$  than if the HK parametrization  is used. HK can
reach quite high maximum mass values, of the order of 2
  $M_\odot$,  for either large values of the vector interaction
even with low or zero magnetic fields or for high magnetic fields and any value
of the vector interaction. Concerning the radii, some comments are in order:
in \cite{Hebeler}, the radii of the canonical $1.4\,M_\odot$ neutron star 
was estimated to lie in the range 9.7-13.9 Km. More recently, there was
a prediction that they should lie in the range $R=9.1^{+1,3}_{-1.5}$ Km
\cite{guillot} and another one stating that the range should be
$10-13.1$ Km~\cite{Lattimer2013}. From Fig. \ref{figs4}, one can see that
there is a window of values for $G_V$ and $B$ which result in radii accepted
by any of the above mentioned analyzes.

\begin{figure}[ht]
\begin{tabular}{cc}
\includegraphics[width=1\linewidth]{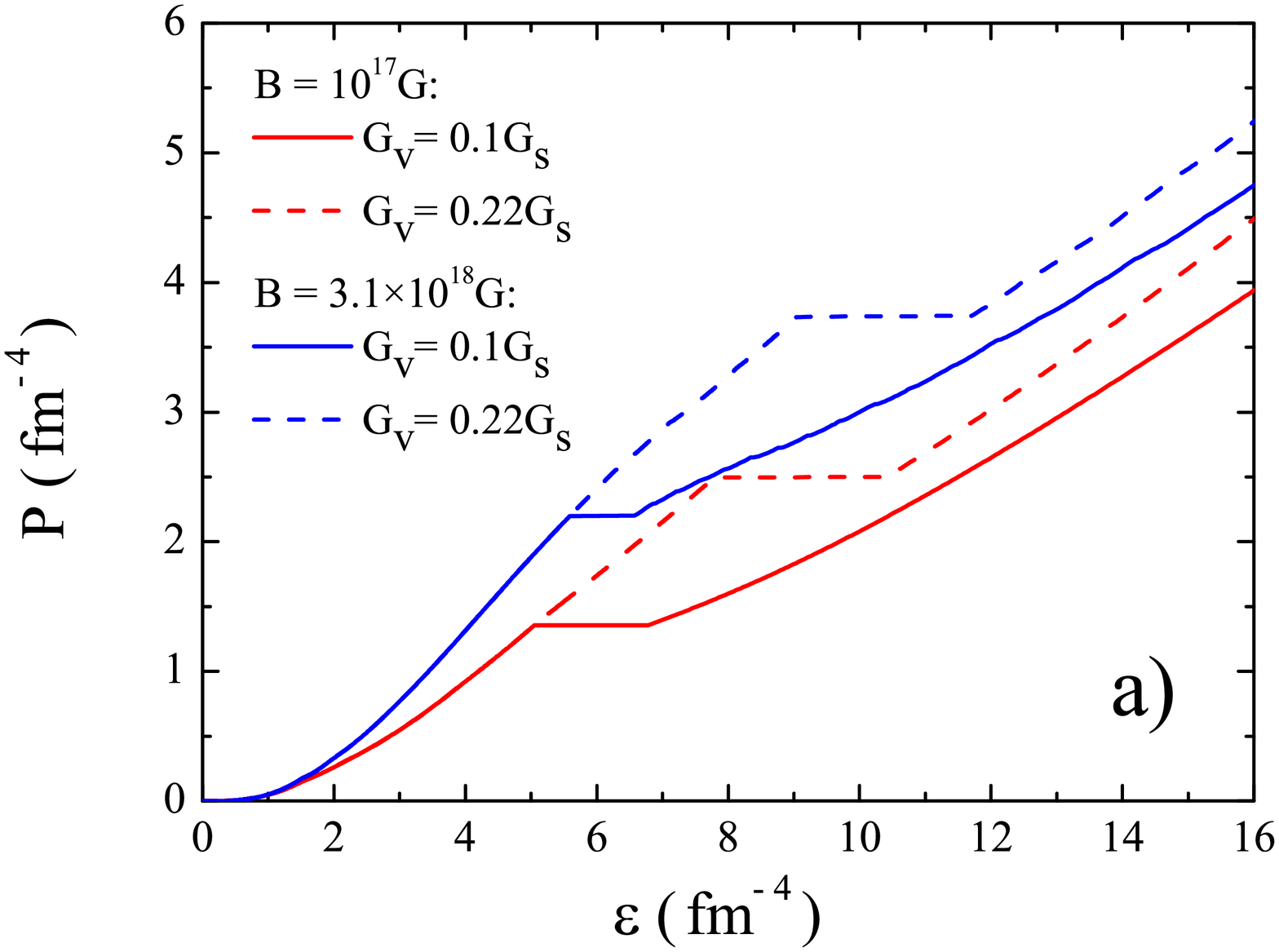}\\
\includegraphics[width=1\linewidth]{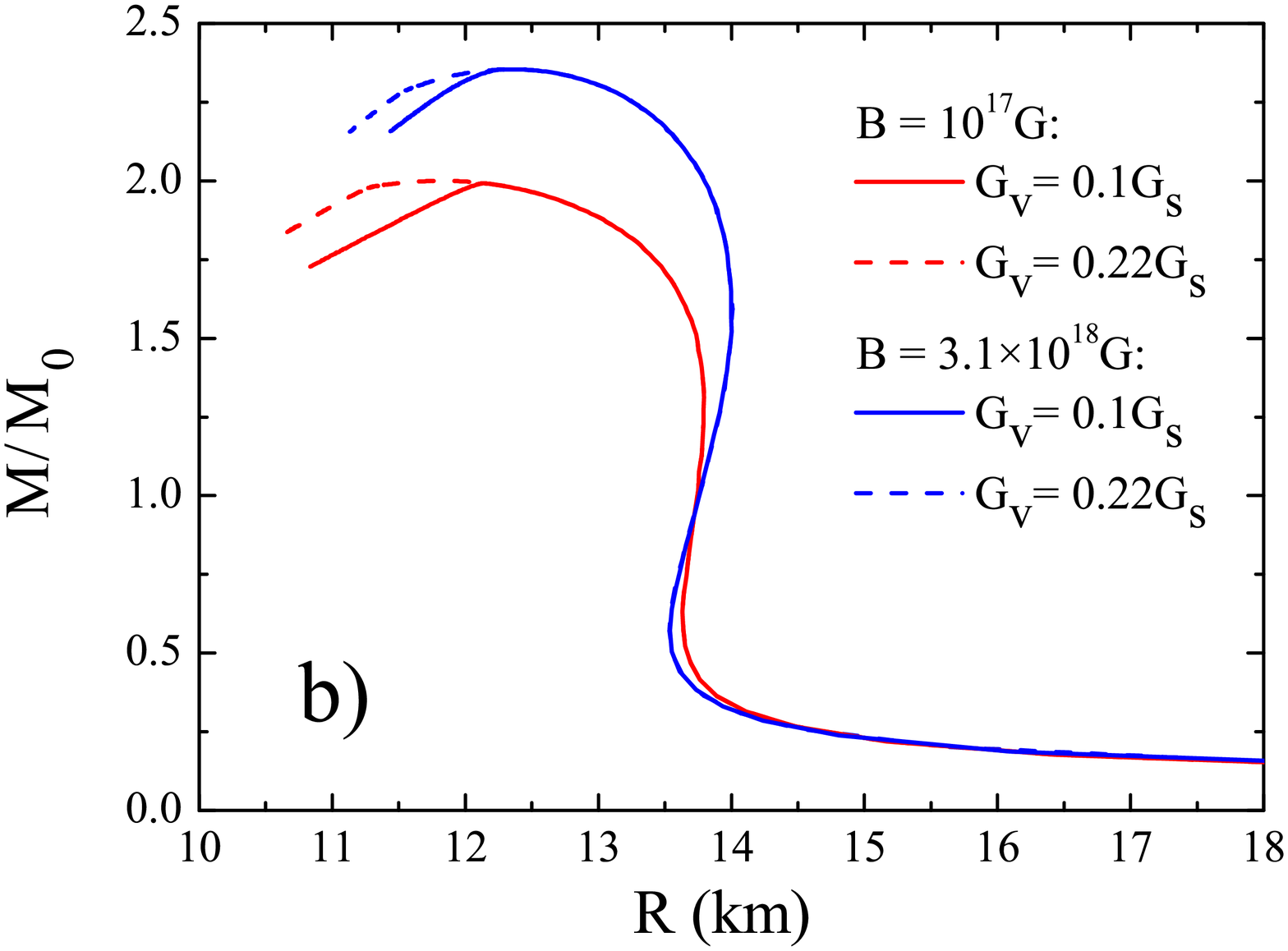}\\
\end{tabular}
\caption{Hybrid star - a) EOS and b)
Mass radius curves obtained with model 2 (P2) for different 
values of $G_V$, two intensities of the magnetic field ( 
$B=10^{17}$ G and $B=3.1 \times 10^{18}$ G) and
parametrization HK.}
\label{figs5}
\end{figure}

\subsection{Stellar matter: hybrid stars}

To make our analysis of the vector interaction as broad as possible, we dedicate
this subsection to revisit the case of hybrid stars under the influence of 
strong magnetic fields. We study the structure of hybrid stars based on the 
Maxwell condition (without a mixed phase), where the hadron phase is described 
by the GM1 \cite{GM1} parametrization of the non-linear Walecka model 
\cite{Serot} 
and the quark phase by the NJL model with the inclusion of the
vector interaction as discussed in the previous subsection. As stated in the 
Introduction, hybrid stars have already been extensively discussed for the 
non-magnetized case \cite{pagliara,bonanno,lenzi2012,logoteta,shao2013,
sasaki2013,masuda2013}.
For the possible existence of magnetars that can be described by hybrid stars, 
the reader can refer to \cite{panda} and \cite{hybrid} and we refrain from
writing the mathematical expressions here. 

\begin{table*}[ht]
\begin{tabular}{|c|c|c|c|c|c|c|c|c|c|c|}
 \hline
 \multicolumn{2}{ |c| }{HK} & $M_{max}$ & $M_{b}$ & R & $\varepsilon_{c}$ & $\varepsilon$ (onset) & $\rho_{c}$ & $\rho$ (onset) & $\mu_{B}(\varepsilon_{c})$ & $\mu_{B}$ (onset) \\

 \multicolumn{2}{ |c| }{} & ($M_{0}$) & ($M_{0}$) & (km) & ($\mathrm{fm}^{-4}$) & ($\mathrm{fm}^{-4}$) &  &  & (MeV) & (MeV) \\
\hline $B=10^{17}$~G & $x=0$ & 1.91 & 2.18 & 12.78 & 4.57 & 3.47 & 0.78 & 0.62 & 1360 & 1330 \\
\cline{2-11}
P2 & $x=0.10$ & 1.99 & 2.30 & 12.14 & 6.27 & 5.05 & - & 0.84 & - & 1503 \\
\cline{2-11}
& $x=0.22$ & 2.00 & 2.31 & 11.82 & 5.93 & 7.79 & 0.95 & 1.18 & 1580 & 1726 \\
\hline
 $B=3.1\times10^{18}$~G & $x=0$ & 2.27 & 2.60 & 12.82 & 4.69 & 3.30 & 0.70 & 0.54 & 1324 & 1261 \\
\cline{2-11}
P2 & $x=0.10$ & 2.35 & 2.70 & 12.34 & 5.29 & 5.59 & 0.74 & 0.78 & 1427 & 1453 \\
\cline{2-11}
& $x=0.22$ & 2.35 & 2.70 & 12.35 & 5.27 & 9.03 & 0.74 & 1.18 & 1426 & 1730 \\
\hline
 \hline
 \multicolumn{2}{ |c| }{RKH} & $M_{max}$ & $M_{b}$ & R & $\varepsilon_{c}$ & $\varepsilon$ (onset) & $\rho_{c}$ & $\rho$ (onset) & $\mu_{B}(\varepsilon_{c})$ & $\mu_{B}$ (onset) \\

 \multicolumn{2}{ |c| }{} & ($M_{0}$) & ($M_{0}$) & (km) & ($\mathrm{fm}^{-4}$) & ($\mathrm{fm}^{-4}$) &  &  & (MeV) & (MeV) \\
\hline $B=10^{17}$~G & $x=0$ & 1.97 & 2.26 & 12.48 & 4.29 & 4.28 & - & 0.74 & - & 1422 \\
\cline{2-11}
P2 & $x=0.10$ & 2.00 & 2.31 & 11.91 & 7.51 & 5.67 & - & 0.92 & - & 1557 \\
\cline{2-11}
& $x=0.19$ & 2.00 & 2.31 & 11.83 & 5.91 & 7.83 & 0.95 & 1.18 & 1579 & 1728 \\
\hline
 $B=3.1\times10^{18}$~G & $x=0$ & 2.33 & 2.69 & 12.79 & 4.69 & 4.19 & - & 0.63 & - & 1335 \\
\cline{2-11}
P2 & $x=0.10$ & 2.35 & 2.70 & 12.34 & 5.30 & 6.52 & 0.74 & 0.88 & 1428 & 1531 \\
\cline{2-11}
& $x=0.19$ & 2.35 & 2.70 & 12.34 & 5.30 & 9.05 & 0.74 & 1.18 & 1428 & 1731 \\
\hline

\end{tabular}
\caption{Stellar macroscopic properties obtained from EOS of
  magnetized hybrid stars built with GM1 and SU(3) NJL with HK and RKH
  parametrizations. $M_{max}$ is the maximum gravitational mass, $M_{b}$ is the maximum baryonic mass, $R$ is the star radius, $\varepsilon_{c}$ is the star central energy density, $\mu_{B}(\varepsilon_{c})$ is the chemical potential for neutron at $\varepsilon_{c}$ and $\mu_{B}$(onset) is the baryonic chemical potential at the onset of the quark phase. }\label{tableh}
\end{table*}

In face of the results we have obtained for quark stars, we next choose to 
construct hybrid stars with the P2 model and both HK and RKH parameter sets 
because this vector interaction term yields the  hardest quark matter EOS. For 
the hadronic phase, we
use the GM1 parametrization \cite{GM1} and hyperon meson coupling constants 
equal to fractions of those of the nucleons, so that $g_{iH}=X_{iH} g_{iN}$, 
where the values of $X_{iH}$ are chosen as $X_{\sigma H}=0.700$ 
and $X_{\omega H}=X_{\rho H}=0.783$ \cite{glen}. This is the same choice
as in \cite{hybrid} for the case of hybrid stars with the quark phase 
described by the NJL model (without the vector interaction).
The EOS obtained with a Maxwell construction
for magnetic fields equal to $B=10^{17}$ G and $B=3.1 \times 10^{18}$ G are
shown in Fig.\ref{figs5}a) for two values of $x$, being $x=0.22$ the maximum
possible value for which a hybrid star can be built with parameter set
HK. For values larger than
0.22, the quark matter EOS becomes too hard and in a pressure versus 
baryonic chemical potential, the hadronic and quark EOS no longer cross each 
other. For an EOS built with GM1 and RKH, the curves are very similar, but
the maximum possible value of $x$ for the crossing of the hadronic and the
quark EOS is 0.19.

Taking into account that NJL does not describe the confinement
  feature of QCD, we cannot, in fact, fix the low-density
  normalization of the pressure. In order to  account for this
  uncertainty the authors of \cite{pagliara,lenzi2012,bonanno,logoteta} have
  included an extra bag pressure that allows the density at
  which  the transition to deconfinement occurs vary. Including this term
  in such a way that the deconfinement transition occurs at lower
  densities than the ones obtained in the present study would have
  allowed us to choose a larger $G_V$ and therefore, a larger maximum
  mass would be possible. In the present study we renormalize the pressure 
in such a way that it is zero for zero baryonic density and do not discuss 
the effect of including an extra bag pressure.

In Fig. \ref{figs5}b) the mass radius curves obtained for the HK 
parametrization from the solution
of the TOV equations are displayed. These macroscopic results are also shown
in Table \ref{tableh}.  In this table we present results for both the
HK and RKH parametrizations, three values of the vector couplings, $x= 0,\,
0.1,$ and the maximum possible value of $x$ for each parameter set,
and two values of the magnetic field intensity
$B=10^{17}$ and $3.1\times 10^{18}$ G. Some of the entrances for the
central baryonic density  are not indicated
because they lie on an intermediate value between the density of the
hadronic phase at the quark phase onset and the corresponding density
of the quark phase.
The only maximum mass configuration that really has a
quark core is obtained for $B=10^{17}$ G and $G_V=0$ within the HK
parametrization, giving rise to a 1.91 $M_\odot$ star. 
It is worth pointing out that the largest maximum masses are now obtained, 
in general, with the parameter set RKH and not HK,  the case of quark 
stars. This is due to the fact that the quark phase sets in at smaller 
densities for the HK parametrization making the EOS softer. 
This result had already been obtained in \cite{hybrid,paoli}.

One can see that the maximum stellar masses 
depend very little on the vector interaction strength. 
 For the larger magnetic field considered, the onset of quark matter
occurs at a larger density than the central density of the maximum
mass hadronic star configuration, for both parametrizations. The same occurs 
for $B=10^{17}$ G and $G_V=0.22$ ($G_V=0.19$)
for the HK (RKH) parameter set.  In these cases the properties of the quark 
phase do not affect the star properties.  On the order hand,  
from Fig.  \ref{figs5}b) for $B=10^{17}$ G and $G_V=0.1$,  it seems that  
as soon as the quark phase sets in the star becomes unstable.
Nevertheless, if we compare the baryonic density at the centre
of the star with the baryonic density at the onset of quarks, we conclude 
that this maximum mass star could, in principle, contain a quark
core.  Had we performed a Gibbs construction, the star core
would be in a mixed phase. All other stars are ordinary hadronic
stars. 

As an overall conclusion, it may be stated that a star that is
  subject to a strong magnetic field attains a smaller baryonic
  density in its centre, and, therefore,  the quark phase is not favored.
 This same conclusion was
  obtained in \cite{panda} where the quark phase was described within
  the MIT bag model. Moreover,  since the inclusion of a
  vector interaction makes the quark  EOS harder, it is also natural to
  expect that a quark EOS with a large $G_V$ difficults the
  occurrence of a quark core. The weak point of the standard NJL model is the
  fact that it does not include confinement, and, therefore, the
  normalization considered for the pressure is not well defined.

Stars with very high masses
are predicted and maximum masses of observed compact stars may set an 
upper limit for the largest possible magnetic field at the centre of the 
star, ∼$2 \times 10^{18}$ G for 2$M_\odot$ stars.

Of course, had we chosen the P1 model to build the hybrid star, the $x$ value 
that would allow for a Maxwell construction would certainly be larger than 
0.19 or 0.22, depending on the choice of parameters, but the stellar maximum 
mass would probably be smaller than 2 $M_\odot$. It is worth remembering that
all results presented here depend also on the choice of the coupling constants 
and meson-hyperon parameters for the hadron phase.

\section{Final remarks}

In the present work we have studied quark matter in the presence of a
strong magnetic field. We had as our main objective understanding the
interplay between the effects of an external magnetic field effect and
the presence of vector interaction in the quark density Lagrangian.
Quark matter was described  within the SU(3) NJL model,  and we have
considered two  forms of the vector interaction, a flavor dependent and 
a flavor independent ones, both frequently used in the literature. 

Two scenarios of homogeneous quark matter were considered: equal flavor 
chemical potential and equal flavor density. In the first scenario the role 
of the vector interaction is an important ingredient,
affecting the fraction of each kind of quarks. 
For the flavor dependent vector interaction the
$s$ quark  fraction increases with the vector interaction.
Within the flavor independent vector interaction the
strangeness sets in at a quite high baryonic density independently of
the vector coupling.  As the presence of $s$ quarks softens the EOS, 
the hardest EOS were obtained with the flavor independent vector interaction. 
The larger the vector coupling the harder the EOS. At low densities the
magnetic field has an effect contrary to the vector interaction and
softens the EOS due to appearance of Landau levels with a large
degeneracy. In fact, if the vector interaction is strong enough the
low density first order transition disappears and a crossover
occurs. The magnetic field, however, increases the range of densities
for which matter is unstable.  On the other hand, at large densities, 
both the vector interaction and the magnetic field act in the same direction, 
in particular, they make the EOS harder. 

Stellar matter and compact star properties in the presence of a static
magnetic field that increases with the baryonic density, have also
been studied.  For quark stars we have shown that the larger the
vector coupling the larger the maximum star mass, independently of the
form of the vector interaction. Moreover, it was shown that the flavor
independent vector interaction predicts larger mass stars, which can
be 0.1- 0.3 $M_\odot$ larger depending on the magnitude of the vector
interaction. The presence of a static  magnetic field increases
the maximum mass, and  masses above $ \sim 2M_\odot$  are obtained for a
magnetic field that is $\sim 3\times 10^{18}$ G in the centre of
the star.

We have shown that within the present quark model hybrid stars with a
quark content in its centre are only possible if  neither  the vector coupling
nor the magnetic fields are too strong.  Strong magnetic fields
disfavor the formation of a quark phase. This fact, however, may have
interesting consequences as already discussed in \cite{panda}, giving
rise to a phase transition when the magnetic field  decays. This kind
of phase transitions are expected to 
release a large amount of energy, possibly in the form of a $\gamma$-ray burst.

\section{Acknowledgements}
This work was partially supported by CNPq (Brazil), CAPES (Brazil) and 
FAPESC (Brazil) under project 2716/2012,TR 2012000344, 
by COMPETE/FEDER and FCT (Portugal) under Grant No. PTDC/FIS/113292/2009 
and by NEW COMPSTAR, a COST initiative.

\end{document}